\documentclass[conference,a4paper]{IEEEtran}
\usepackage{xcolor}
\usepackage{balance}
\usepackage{cite}
\usepackage{multirow}
\usepackage[pdftex]{graphicx}
\usepackage{amsmath,amssymb,amsfonts}
\usepackage{textcomp}
\usepackage[caption=false,font=footnotesize]{subfig}
\usepackage[nolist]{acronym}
\usepackage[capitalize]{cleveref}
\usepackage{nicefrac}
\usepackage{textcomp}
\usepackage{graphicx}  
\crefname{table}{Tab.}{Tabs.}
\Crefname{table}{Tab.}{Tabs.}
\crefformat{equation}{(#2#1#3)}

\renewcommand{\arraystretch}{1.4}

\newcommand{\TRx}[3]{#1_{\rm #2}^{\left( {\rm #3} \right)} }

\begin{document}
\begin{acronym}
\acro{AWG}[AWG]{Arbitrary Waveform Generator}
\acro{CW}[CW]{Continuous Wave}
\acro{ADC}[ADC]{analog-to-digital converter}
\acro{DAC}[DAC]{digital-to-analog converter}
\acro{HST}[HST]{High-Speed Train}
\acro{IF}[IF]{Intermediate Frequency}
\acro{LO}[LO]{local oscillator}
\acro{LSF}[LSF]{local scattering function}
\acro{MMW}[mmWave]{millimeter wave}
\acro{OFDM}[OFDM]{orthogonal frequency-division multiplexing}
\acro{PCB}[PCB]{Printed Circuit Board}
\acro{SMD}[SMD]{Surface Mount Device}
\acro{SNR}[SNR]{signal-to-noise ratio}
\acro{SINR}[SINR]{signal-to-interference-and-noise ratio}
\acro{RF}[RF]{radio frequency}
\acro{V2X}[V2X]{vehicle-to-everything}
\acro{IC}[IC]{Integrated Circuit}
\acro{FPGA}[FPGA]{Field Programmable Gate Array}
\acro{ISI}[ISI]{Inter-Symbol Interference}
\acro{OOBA-MRC}[OOBA-MRC]{out-of-band aided maximal ratio combining}
\acro{CS}[CS]{compressed sensing}
\acro{CSIT}{channel state information at the transmitter}
\acro{CSI}{channel state information}
\acro{ML}{machine learning}
\acro{IL}{insertion loss}
\acro{RNN}{recurrent neural networks}
\acro{EVM}{error vector magnitude}
\acro{OMP}{orthogonal matching pursuit}
\acro{PA}[PA]{power amplifier}
\acro{LNA}[LNA]{low noise amplifier}
\acro{EIRP}{effective isotropic radiated power}
\acro{LSTM}{long short-term memory}
\acro{GRU}{Gated Recurrent Unit}
\acro{FDD}{frequency-division duplex}
\acro{TDD}{time-division duplex}
\acro{CIR}{channel impulse response}
\acro{CTF}{channel transfer function}
\acro{PDP}{power delay profile}
\acro{DSD}{Doppler power spectral density}
\acro{IFFT}{Inverse Fast Fourier Transform}
\acro{ITS}[ITS]{intelligent transportation systems}
\acro{5G}[5G]{fifth generation}
\acro{NR}[NR]{new radio}
\acro{QAM}[QAM]{quadrature amplitude modulation}
\acro{ICI}[ICI]{Inter-Carrier Interference}
\acro{MSE}[MSE]{mean squared error}
\acro{BER}[BER]{bit error ratio}
\acro{RMS}[RMS]{root-mean-square}
\acro{TDL-A}[TDL-A]{tapped delay line A}
\acro{TDL-D}[TDL-D]{tapped delay line D}
\acro{ISI}[ISI]{Inter-Symbol Interference}
\acro{DFT}[DFT]{discrete Fourier transform}
\acro{IDFT}[IDFT]{inverse discrete Fourier transform}
\acro{CTF}[CTF]{channel transfer function}
\acro{DR}[DR]{dynamic range}
\acro{HPBW}[HPBW]{half-power beamwidth}
\acro{DPSS}[DPSS]{discrete prolate spheroidal sequences}
\acro{CDF}[CDF]{cumulative distribution function}
\acro{URLLC}[URLLC]{ultra-reliable low-latency communication}
\acro{3GPP}[3GPP]{3rd Generation Partnership Project}
\acro{MIMO}[MIMO]{multiple-input multiple-output}
\acro{SISO}[SISO]{single-input single-output}
\acro{MRC}[MRC]{maximal ratio combining}
\acro{AWGN}[AWGN]{additive white Gaussian noise}
\acro{LS}[LS]{least-squares}
\acro{SVD}[SVD]{singular value decomposition}
\acro{SE}[SE]{spectral efficiency}
\acro{ULA}[ULA]{uniform linear array}
\acro{UPA}[UPA]{uniform planar array}
\acro{FSPL}[FSPL]{free space path loss}
\acro{LOS}[LOS]{line-of-sight}
\acro{NLOS}[NLOS]{non-line-of-sight}
\acro{AoA}[AoA]{angle-of-arrival}
\acro{MUSIC}[MUSIC]{multiple signal classification}
\acro{EVD}[EVD]{eigenvalue decomposition}
\acro{AoD}[AoD]{angle-of-departure}
\acro{RMSE}[RMSE]{root mean squared error}
\acro{SIMO}[SIMO]{single-input multiple-output}
\acro{MAC}[MAC]{medium access control}
\acro{eMBB}[eMBB]{enhanced mobile broadband}
\acro{SVM}[SVM]{support vector machine}
\acro{VNA}[VNA]{vector network analyzer}
\acro{CNN}[CNN]{convolutional neural network}
\acro{DNN}[DNN]{deep neural network}
\acro{EE}[EE]{energy efficiency}
\acro{ReLU}[ReLU]{rectified linear unit}
\acro{ZP}[ZP]{zero padding}
\acro{BN}[BN]{batch normalization}
\acro{NN}[NN]{neural network}
\acro{ADAM}[ADAM]{adaptive moment estimation}
\acro{AQNM}[AQNM]{additive quantization noise model}
\end{acronym}
\title{Multi-Band Patch Antenna Array for Out-of-Band Aided Millimeter Wave Communication}

\author{\IEEEauthorblockN{
Faruk Pasic\IEEEauthorrefmark{1},
Jure Soklič\IEEEauthorrefmark{2},
Robert Langwieser\IEEEauthorrefmark{1},
Stefan Schwarz\IEEEauthorrefmark{1} and
Christoph F. Mecklenbräuker\IEEEauthorrefmark{1}
}%

\IEEEauthorblockA{\IEEEauthorrefmark{1}
Institute of Telecommunications, TU Wien, Vienna, Austria}
\IEEEauthorblockA{\IEEEauthorrefmark{2}
PIDSO Propagation Ideas and Solutions GmbH, Vienna, Austria}
\IEEEauthorblockA{faruk.pasic@tuwien.ac.at}
}

\IEEEoverridecommandlockouts 

\makeatletter
\def\thanks#1{\protected@xdef\@thanks{\@thanks
        \protect\footnotetext{#1}}}
\makeatother

\maketitle

\begin{abstract}
Future wireless communication systems will integrate both sub-6\,GHz and \ac{MMW} frequency bands within multi-antenna architectures to meet the increasing demand for high data rates.
In such multi-band systems, reliable information obtained from the sub-6\,GHz band can be exploited to support communication at \ac{MMW} frequencies.
To ensure that both systems experience similar multi-path propagation effects, the sub-6\,GHz and \ac{MMW} antenna arrays have to be co-located and precisely aligned.
However, such a configuration may adversely alter the radiation characteristics of the arrays, potentially degrading their performance.
In this paper, we investigate the impact of positioning a \ac{MMW} antenna structure in front of a sub-6\,GHz antenna structure. 
Through both simulations and measurements, we evaluate how the presence of the \ac{MMW} structure affects the radiation pattern of the sub-6\,GHz one.
The results demonstrate that the influence of the \ac{MMW} structure on the sub-6\,GHz performance is minor, indicating that co-located configurations are feasible with negligible degradation.
\end{abstract}
\vskip0.5\baselineskip

\acresetall

\section{Introduction}
Due to the high utilization of conventional sub-6\,GHz frequency bands, the available bandwidth is limited, which constrains data transmission rates.
In contrast, \ac{MMW} frequency bands (24\,GHz -- 300\,GHz)~\cite{3gpp.38.101-1} offer significantly wider bandwidths, making them well-suited to support the growing demand for high data rates~\cite{Molisch2025}.
Furthermore, \ac{MMW} systems are increasingly being deployed in conjunction with sub-6\,GHz systems to enable multi-band communication and improve reliability~\cite{Shafi2020, Hofer2025}.

Compared to \ac{MMW} bands, which suffer from high propagation losses, sub-6\,GHz bands exhibit more favorable propagation characteristics, resulting in more robust link reliability.
This makes them a valuable source of out-of-band information that can be leveraged to support communication at \ac{MMW} frequencies.
So far, numerous methods that exploit sub-6\,GHz out-of-band information to aid \ac{MMW} communication have been proposed~\cite{Ali2018, Pasic2024_TCOM, Ali2019, Pasic2025_infocom}. 
These methods employ co-located and aligned sub-6\,GHz and \ac{MMW} antenna arrays (sharing the same central point) to ensure that both systems experience similar multi-path propagation effects.

However, the design of such co-located and aligned multi-band antenna arrays poses significant practical challenges. 
A common implementation involves placing the \ac{MMW} array in front of the sub-6\,GHz array.
While this configuration allows for spatial alignment, it can adversely affect the radiation pattern of the sub-6\,GHz array, potentially degrading its performance.
Therefore, it is crucial to evaluate the performance of such multi-band antenna structures under realistic hardware constraints.

\begin{figure}
	\centering
	\includegraphics[width=0.85\columnwidth]{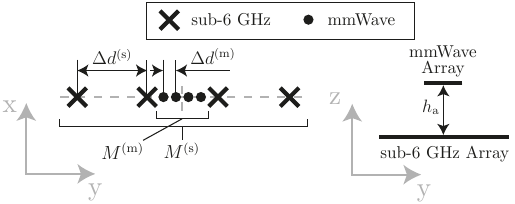}
	\caption{The multi-band uniform linear array (ULA) consists of sub-6\,GHz and mmWave antenna arrays that are co-located and precisely aligned (sharing the same geometric center) and separated by an inter-array spacing of $h_{\rm a}$.}
	\label{fig:antenna_geometry}
\end{figure}

\textbf{Contribution:}
In this paper, we investigate the impact of positioning a \ac{MMW} antenna structure in front of a sub-6\,GHz structure. 
We first assess, through both simulations and measurements, how the presence of the \ac{MMW} array influences the radiation pattern of a single sub-6\,GHz element. 
The analysis is then extended to full sub-6\,GHz and \ac{MMW} arrays using simulations.

\textbf{Organization:}
\cref{sec:array_design} describes the multi-band antenna array design used in this paper.
Simulation and measurement results are presented in~\cref{sec:results}.
Finally, \cref{sec:conclusion} concludes the paper.

\textbf{Notation:}
The superscript $\left( \cdot \right) ^{\left( \rm b \right)}$ stands for frequency-band dependent values, where ${\rm b} \in \{ {\rm s}, {\rm m} \}$.
In this context, ${\rm s}$ stands for the sub-6\,GHz frequency band and ${\rm m}$ stands for the \ac{MMW} frequency band.

\section{Multi-Band Antenna Array Design} \label{sec:array_design}
\begin{figure}
	\centering
	\includegraphics[width=\columnwidth]{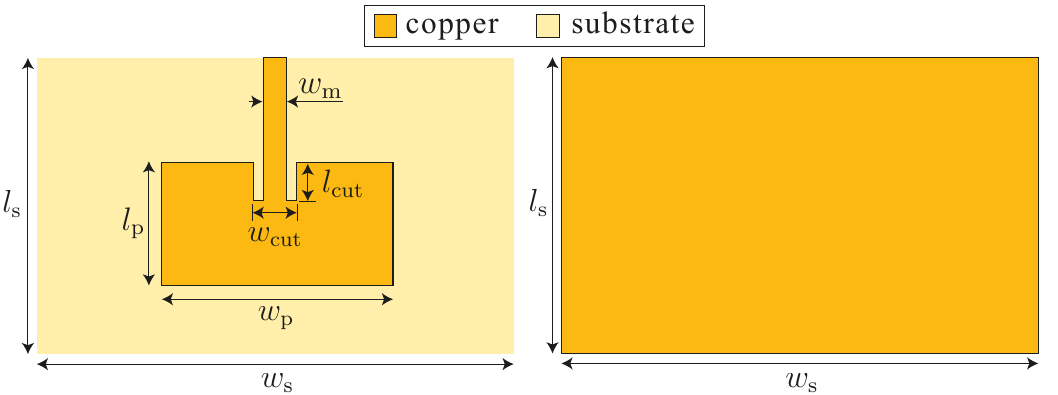}
	\caption{On the top layer (left), each patch element, with dimensions $\TRx{l}{p}{b}\times\TRx{w}{p}{b}$, is fed by a microstrip line of width $\TRx{w}{m}{b}$ and features an inset feed of dimensions $\TRx{l}{cut}{b}\times\TRx{w}{cut}{b}$. The bottom layer (right) consists of a solid copper ground plane.}
	\label{fig:patch_element}
\end{figure}

We consider a multi-band antenna array composed of co-located sub-6\,GHz and \ac{MMW} arrays, operating simultaneously in the radiative far-field regime.
The sub-6\,GHz array comprises $\TRx{M}{}{s}$ elements, while the \ac{MMW} array consists of $\TRx{M}{}{m}$ elements (see~\cref{fig:antenna_geometry}).
The arrays are precisely aligned, sharing the same geometric center, with an inter-array spacing of $h_{\rm a}$, as illustrated in~\cref{fig:antenna_geometry}.
The sub-6\,GHz and \ac{MMW} arrays are implemented as \acp{ULA} of patch antenna elements. 
In both arrays, adjacent antenna elements are mutually separated by $\Delta \TRx{d}{}{b} = \,0.5\,\TRx{\lambda}{}{b}$, where $\TRx{\lambda}{}{b}$ denotes the wavelength corresponding to the respective frequency band.
Each patch antenna element has dimensions $\TRx{l}{p}{b}\times\TRx{w}{p}{b}$ and is fed by a microstrip line of width $\TRx{w}{m}{b}$ with inset feed dimensions $\TRx{l}{cut}{b}$ and $\TRx{w}{cut}{b}$.
\cref{fig:patch_element} shows top and bottom layers of a single patch antenna.
The patch elements are designed to resonate at the carrier frequency $\TRx{f}{c}{b}$ of their respective systems.
The patch elements are etched on a substrate of thickness $\TRx{h}{s}{b}$ with copper thickness $\TRx{h}{c}{b}$, relative dielectric constant $\TRx{\varepsilon}{r}{b}$ and dissipation factor $\TRx{\tan\delta}{}{b}$.
The total array dimensions are given by $\TRx{l}{s}{b}\times\TRx{w}{s}{b}$, where $\TRx{l}{s}{b} = \Delta \TRx{d}{}{b} + \TRx{\lambda}{}{b}/2$ and $\TRx{w}{s}{b} = \TRx{M}{}{b} \Delta \TRx{d}{}{b} + \TRx{\lambda}{}{b}/2.$
The specific parameter values for each frequency band are summarized in~\cref{tab:simParams}.

\begin{table}[t]
	\centering
	\caption{Simulation Parameters} 
	\label{tab:simParams}
	\begin{tabular}{rcc}
		\hline
		\textbf{Parameter}                          & \multicolumn{2}{c}{\textbf{Value}} \\ \hline
		Frequency Band                              & sub-6 GHz         & mmWave         \\
		Carrier Frequency $f_{\rm c}$               & 2.55\,GHz         & 25.5\,GHz       \\
		Wavelength $\lambda$                        & 11.76\,cm         & 1.176\,cm       \\
		Substrate Type                              & RT-Duroid 5880    & RO4350B     \\
		Relative Dielectric Constant $\varepsilon_{\rm r}$ & 2.2        & 3.66 \\
		Substrate Thickness $h_{\rm s}$             & 1.5\,mm           & 0.254\,mm     \\
		Copper Thickness $h_{\rm c}$                & 70\,\textmu m    & 70\,\textmu m   \\
		Dissipation Factor $\tan\delta$             & 0.0004            & 0.0037 \\
		Microstrip Line Width $w_{\rm m}$           & 4.53307\,mm       & 0.504685\,mm  \\
		Patch Element Width $w_{\rm p}$             & 46.47\,mm         & 3.851\,mm \\
		Patch Element Length $l_{\rm p}$            & 38.89\,mm         & 2.985\,mm \\
		Inset Feed Cut Width $w_{\rm cut}$          & 5.7899\,mm        & 0.5276\,mm \\
		Inset Feed Cut Length $l_{\rm cut}$         & 10\,mm            & 1.005\,mm \\        
		\hline
	\end{tabular}
\end{table}

Moreover, we validate the accuracy of our antenna design in terms of return loss through simulations in ANSYS HFSS~\cite{HFSS}.
For the sub-6\,GHz array, the antenna resonates at 2.55\,GHz with a return loss of 18.47\,dB, whereas the \ac{MMW} array resonates at 25.5\,GHz with a return loss of 13.16\,dB.
These results demonstrate satisfactory impedance matching at the target frequencies in both frequency bands.
Furthermore, we complement the simulations results with experimental measurements.
A single sub-6\,GHz patch element is manufactured in-house according to the design parameters listed in~\cref{tab:simParams}.
We measure the return loss of the fabricated patch antenna using a \ac{VNA} (Rohde\&Schwarz ZVA8) under controlled laboratory conditions.
The measured results show that the antenna resonates at 2.5625\,GHz with a return loss of 18.22\,dB, corresponding to only a 0.5\% deviation from the target frequency of 2.55\,GHz.
This small shift can be attributed to fabrication tolerances and substrate parameter variations, which are common in practical antenna prototyping.
At the exact target frequency of 2.55\,GHz, the measured return loss is 12.96\,dB, which still indicates satisfying impedance matching and acceptable performance for practical operation.

\section{Simulation and Measurement Results} \label{sec:results}
In this section, we investigate the effect of placing a \ac{MMW} antenna structure in front of a sub-6\,GHz antenna structure.
\cref{subsec:single_element_results} focuses on a single sub-6\,GHz antenna element, where the influence of the \ac{MMW} array is assessed through both simulations and measurements.
\cref{subsec:array_results} extends the analysis to full antenna arrays, considering \acp{ULA} using simulations.
Consistent with common practice, the \ac{MMW} arrays are modeled with a larger number of elements compared to their sub-6\,GHz counterparts~\cite{Heath2016}.

\subsection{Single Antenna Element Analysis} \label{subsec:single_element_results}
We first analyze the impact of placing a \ac{MMW} \ac{ULA} with different dimensions (i.e., number of antenna elements) in front of a single sub-6\,GHz patch antenna element.
Specifically, for the \ac{MMW} array, we consider $\TRx{M}{}{m} > 1$, while for the sub-6\,GHz antenna we have $\TRx{M}{}{s} = 1$.
The analysis is carried out using both full-wave simulations and antenna measurements.

\begin{figure}[t]
	\centering
	\includegraphics[width=0.9\columnwidth]{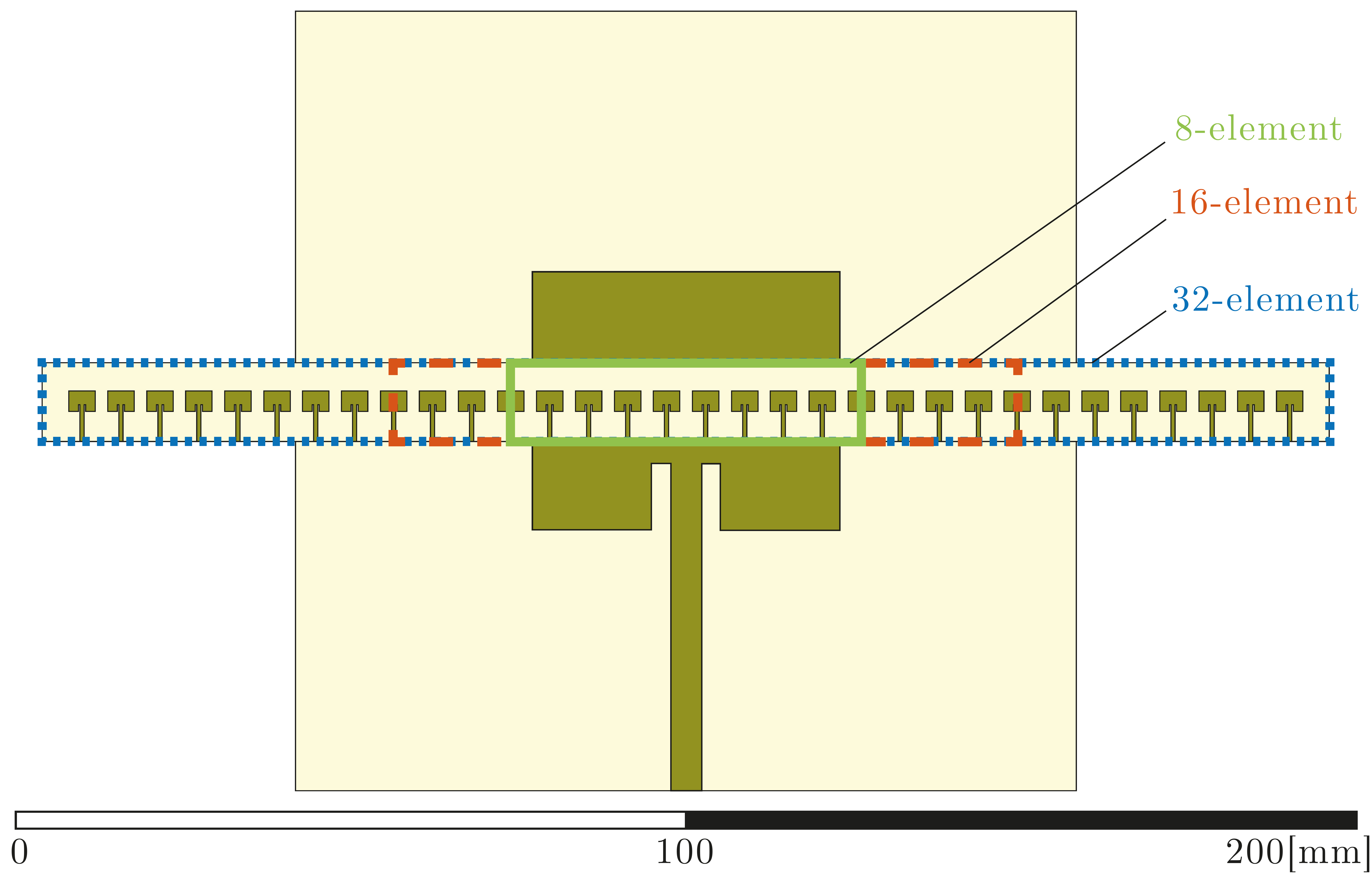}
	\caption{A \ac{MMW} \ac{ULA} with variable size (8-element, 16-element or 32-element) is positioned in front of the single sub-6\,GHz patch antenna.}
	\label{fig:single_patch_ULA}
\end{figure}

\begin{figure}[t]
    \centering
    \includegraphics[width=\columnwidth]{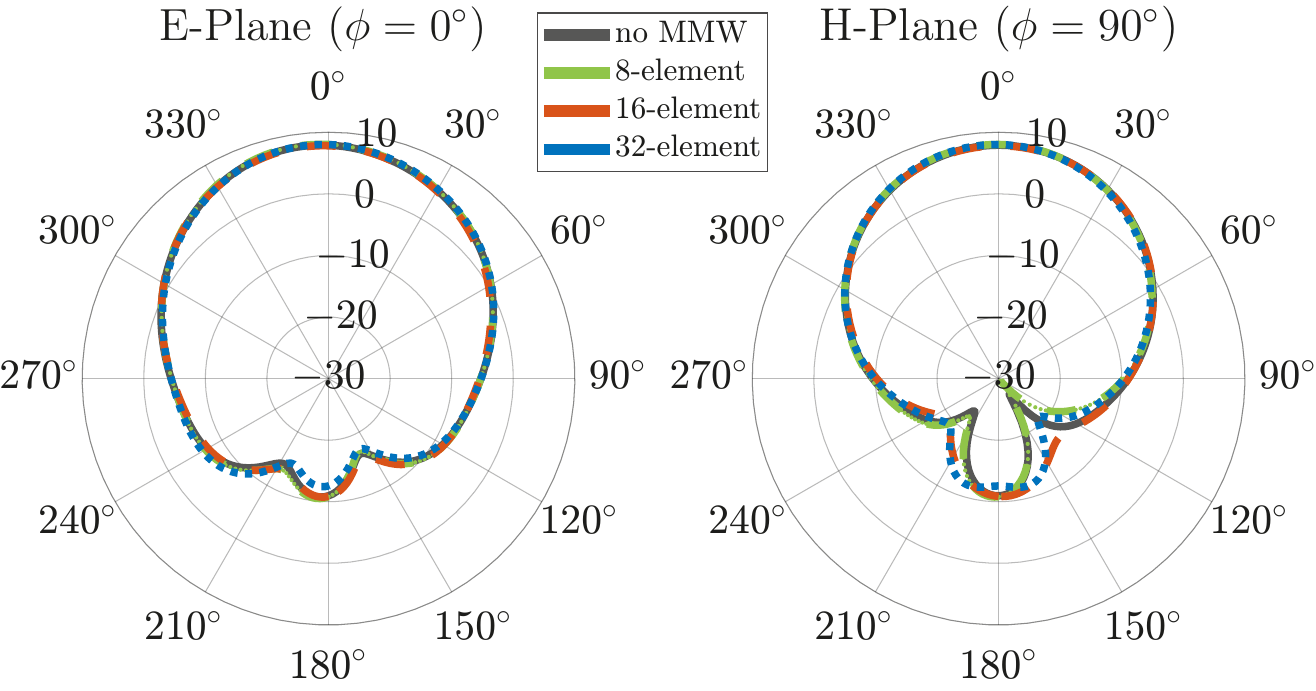}
	\caption{The realized gain of the sub-6\,GHz patch element in the main lobe direction is approximately 7.96\,dBi, with a negligible degradation of around 0.12\,dB when the \ac{MMW} array is present.}
	\label{fig:single_gain_conf}
\end{figure}

The simulations are performed at 2.55\,GHz using the commercial software ANSYS HFSS~\cite{HFSS}.
The simulated structure with an inter-array spacing of $h_{\rm a}=10$\,mm is shown in~\cref{fig:single_patch_ULA}.
We show in~\cref{fig:single_gain_conf} the realized gain patterns of the sub-6\,GHz element for different \ac{MMW} array configurations in the E-plane and H-plane.
The results for the realized gain in the main-lobe direction and the \ac{HPBW} are summarized in~\cref{tab:results}.
As a baseline reference, we use the realized gain of the standalone sub-6\,GHz patch element without any \ac{MMW} structure present, which is approximately 7.96\,dBi. 

\renewcommand{\arraystretch}{1.3}
\begin{table}[t]
	\centering
	\caption{Results} 
	\label{tab:results}
	\begin{tabular}{c c c | c | c c}
		\hline
		  & \multicolumn{1}{c|}{} &  \textbf{mmWave Array}    & \multicolumn{1}{c|}{\textbf{Realized}} & \multicolumn{2}{c}{\textbf{HPBW}}  \\  
        & \multicolumn{1}{c|}{} & \textbf{Configuration}    & \multicolumn{1}{c|}{\textbf{Gain}}     & E-plane       & H-plane \\
		\hline \hline
        \multicolumn{1}{c|}{  \multirow{7}{*}{\rotatebox[origin=c]{90}{Single Antenna}} } & \multicolumn{1}{c|}{ \multirow{4}{*}{\rotatebox[origin=c]{90}{Simulated}} } &
		                                                  no mmWave         & 7.96\,dBi     & 68$^\circ$    & 64$^\circ$   \\
		\multicolumn{1}{c|}{} & \multicolumn{1}{c|}{} & 8-element         & 7.86\,dBi     & 68$^\circ$    & 62$^\circ$   \\
		\multicolumn{1}{c|}{} & \multicolumn{1}{c|}{} & 16-element        & 7.84\,dBi     & 68$^\circ$    & 64$^\circ$   \\
		\multicolumn{1}{c|}{} & \multicolumn{1}{c|}{} & 32-element        & 7.86\,dBi     & 68$^\circ$    & 62$^\circ$   \\
		\cline{2-6}
		\multicolumn{1}{c|}{} & \multicolumn{1}{c|}{ \multirow{3}{*}{\rotatebox[origin=c]{90}{Measured}} } &                             
                                                        no mmWave         & 7.20\,dBi     & 75$^\circ$    & 74$^\circ$   \\
		\multicolumn{1}{c|}{} & \multicolumn{1}{c|}{} & 8-element         & 6.95\,dBi     & 71$^\circ$    & 74$^\circ$   \\
		\multicolumn{1}{c|}{} & \multicolumn{1}{c|}{} & 16-element        & 7.10\,dBi     & 75$^\circ$    & 74$^\circ$   \\
		\hline
		\multicolumn{1}{c|}{  \multirow{4}{*}{\rotatebox[origin=c]{90}{Array}} } & \multicolumn{1}{c|}{ \multirow{4}{*}{\rotatebox[origin=c]{90}{Simulated}} } &
                                                        no mmWave         & 15.33\,dBi    & 68$^\circ$    & 12$^\circ$   \\
		\multicolumn{1}{c|}{} & \multicolumn{1}{c|}{} & 8-element         & 15.23\,dBi    & 68$^\circ$    & 12$^\circ$   \\
		\multicolumn{1}{c|}{} & \multicolumn{1}{c|}{} & 16-element        & 15.20\,dBi    & 66$^\circ$    & 12$^\circ$   \\
		\multicolumn{1}{c|}{} & \multicolumn{1}{c|}{} & 32-element        & 15.18\,dBi    & 68$^\circ$    & 12$^\circ$   \\        
		\hline \hline
	\end{tabular}
\end{table}

The results demonstrate that the presence of the \ac{MMW} array has only a negligible effect on the radiation performance of the sub-6\,GHz patch. 
In particular, the maximum observed degradation in realized gain is around 0.12\,dB.
The \ac{HPBW} in the E-plane remains unchanged at 68$^\circ$, while in the H-plane it decreases slightly from 64$^\circ$ to 62$^\circ$ (a reduction of 3\%).
The results indicate that the electromagnetic interaction between the two structures is minimal at sub-6\,GHz frequencies.

\begin{figure}
	\centering
	\includegraphics[width=0.92\columnwidth]{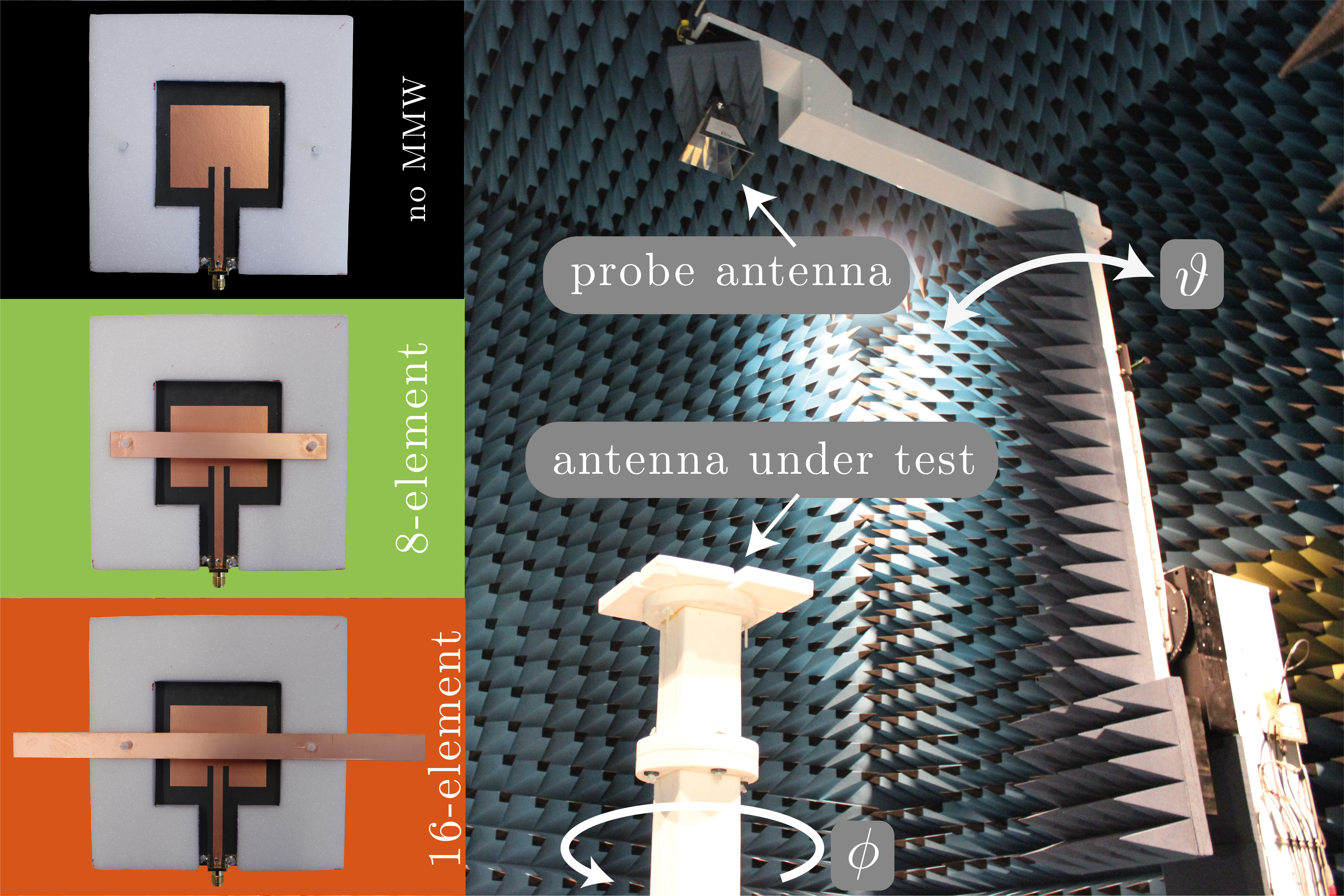}
	\caption{Measurement setup for field-pattern measurements conducted in the anechoic chamber at TU Wien.}
	\label{fig:campaign}
\end{figure}

Furthermore, the simulation results are complemented by field-pattern measurements conducted in the anechoic chamber at TU Wien.
The measurement setup, shown in~\cref{fig:campaign}, employs a horn antenna as the probe antenna and the in-house manufactured sub-6\,GHz patch antenna as the antenna under test.
To emulate the presence of a \ac{MMW} array in front of the patch, unetched printed circuit boards with copper cladding on both sides are used.
These boards, fabricated on the \ac{MMW} substrate with the thickness specified in~\cref{tab:simParams}, have dimensions corresponding to 8-element and 16-element \ac{MMW} arrays. 
The \ac{MMW} printed circuit boards are placed with an inter-array spacing of $h_{\rm a}=10$\,mm from the sub-6\,GHz patch, with a ROHACELL layer in between. 
The realized gain of the single sub-6\,GHz patch element is then measured at 2.55\,GHz.

\begin{figure}[t]
    \centering
    \includegraphics[width=\columnwidth]{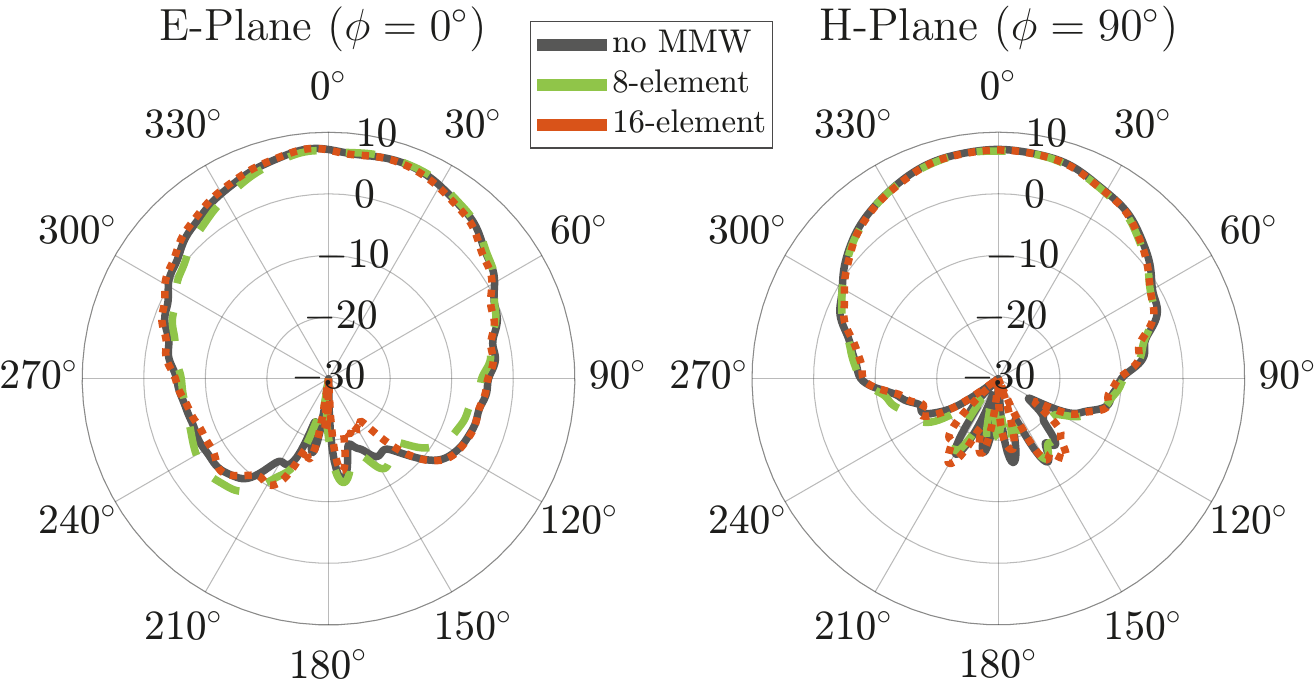}
	\caption{The measured realized gain of the sub-6\,GHz patch element in the main lobe direction is approximately 7.2\,dBi, with a degradation of around 0.25\,dB when the \ac{MMW} array is present.}
	\label{fig:measured_single_gain_conf}
\end{figure}

We show the measured gain patterns of the sub-6\,GHz element for different \ac{MMW} array configurations in the E-plane and H-plane in~\cref{fig:measured_single_gain_conf}.
As a baseline, we measure the realized gain of the standalone sub-6\,GHz patch element without \ac{MMW} printed circuit boards present. 
The realized gain of the baseline configuration is approximately 7.2\,dBi in the main-lobe direction.
The corresponding measured \ac{HPBW} is 75$^\circ$ in the E-plane and 74$^\circ$ in the H-plane.
Consistent with the simulation analysis, the measurements confirm that the presence of the \ac{MMW} array has only a negligible effect on the radiation performance of the sub-6\,GHz patch.
As shown in~\cref{tab:results}, the maximum observed degradation in realized gain is about 0.25\,dB in the main-lobe direction.
The \ac{HPBW} in the E-plane decreases slightly from 75$^\circ$ to 71$^\circ$ (a 5\% reduction), while in the H-plane it remains unchanged at 74$^\circ$.

\subsection{Antenna Array Analysis} \label{subsec:array_results}
\begin{figure}[t]
    \centering
    \includegraphics[width=\columnwidth]{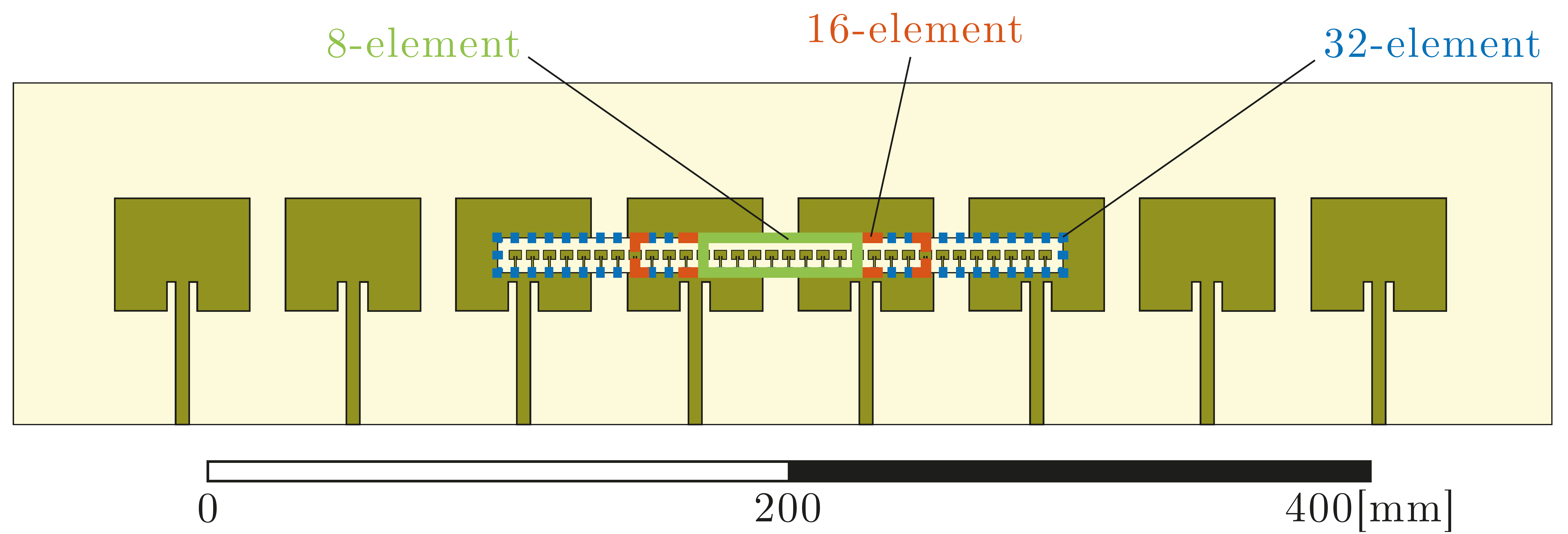}
    \caption{A \ac{MMW} \ac{ULA} with variable size (8-element, 16-element or 32-element) is positioned in front of an 8-element sub-6\,GHz \ac{ULA}.}
    \label{fig:array_patch_ULA}
\end{figure}

\begin{figure}[t]
    \centering
    \includegraphics[width=\columnwidth]{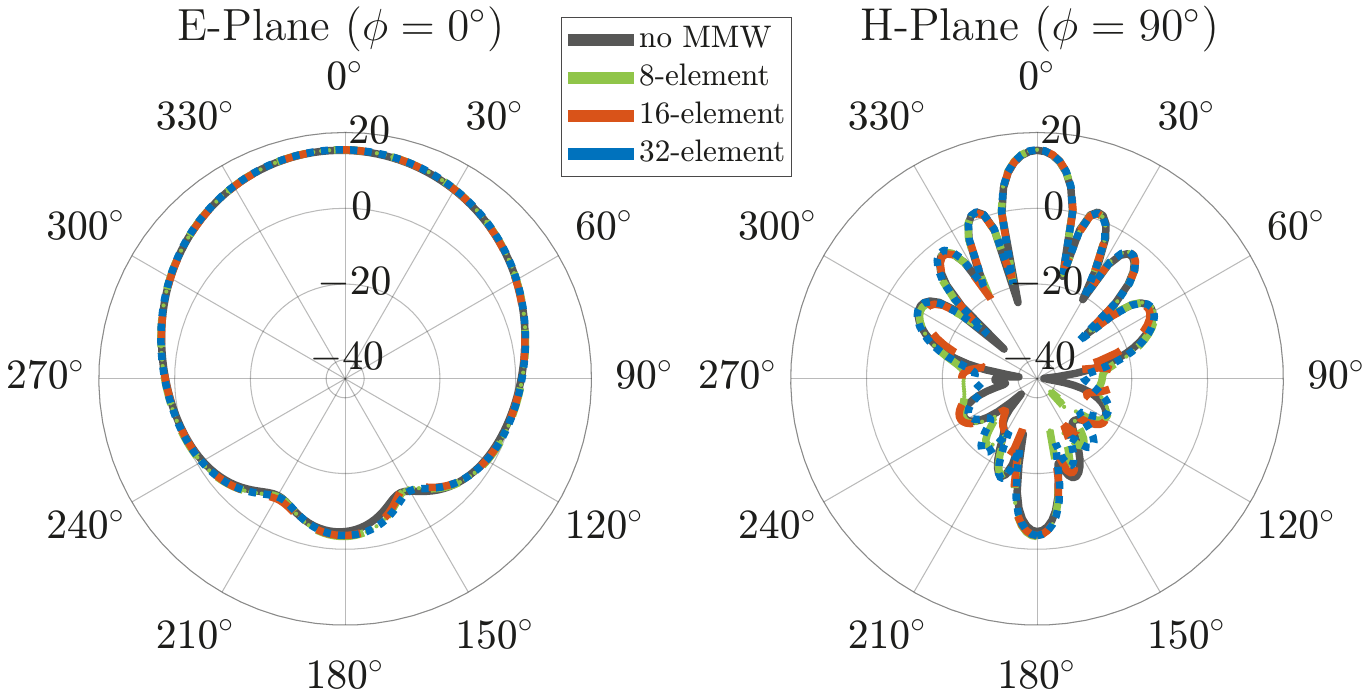}
	\caption{The realized gain of the 8-element sub-6\,GHz \ac{ULA} is approximately 15.33\,dBi in the main lobe direction, with only a negligible reduction of about 0.15\,dB due to the presence of the \ac{MMW} array.}
	\label{fig:ULA_array_gain_conf}
\end{figure}

In this analysis, we investigate the impact of placing \ac{MMW} arrays of varying size in front of sub-6\,GHz arrays through simulations.
We evaluate the radiation performance of the sub-6\,GHz arrays at 2.55\,GHz using the commercial software ANSYS HFSS~\cite{HFSS}. 
The simulated structure, with an inter-array spacing of $h_{\rm a}=10$\,mm, is shown in~\cref{fig:array_patch_ULA}.
In~\cref{fig:ULA_array_gain_conf}, we show the realized gain patterns of the 8-element sub-6\,GHz array in the E-plane and H-plane.

As a baseline, we employ the realized gain of the 8-element sub-6\,GHz \ac{ULA} without the \ac{MMW} structure, which is approximately 15.33\,dBi in the main-lobe direction.
As shown in~\cref{tab:results}, the presence of the \ac{MMW} array introduces only a negligible effect: the maximum reduction in realized gain is about 0.15\,dB in the main-lobe direction,  the E-plane \ac{HPBW} decreases slightly from 68$^\circ$ to 66$^\circ$ (3\%) and the H-plane \ac{HPBW} remains unchanged at 12$^\circ$.

\begin{figure}[t]
    \centering
    \includegraphics[width=\columnwidth]{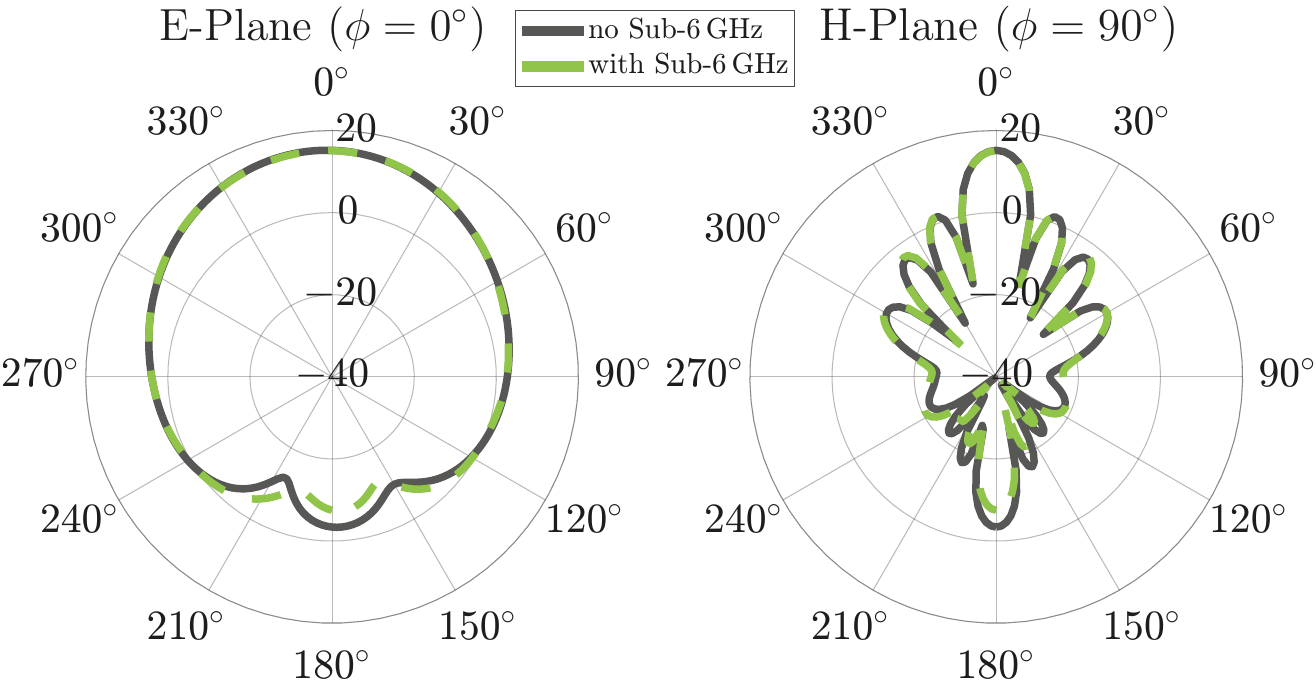}
	\caption{The sub-6\,GHz \ac{ULA} acts as an additional ground plane for the \ac{MMW} \ac{ULA}, resulting in negligible impact on the main-lobe gain while effectively suppressing the back lobe.}
	\label{fig:gain_array_height_MMW}
\end{figure}

Finally, we analyze the reverse scenario, i.e., the effect of a sub-6\,GHz array on the radiation performance of a \ac{MMW} array.
For this case, as shown in~\cref{fig:gain_array_height_MMW}, we simulate the realized gain of an 8-element \ac{MMW} \ac{ULA} at 25.5\,GHz using the HFSS.
The realized gain is only marginally reduced, from 15.16\,dBi in the baseline configuration without the sub-6\,GHz array to 15.08\,dBi when the sub-6\,GHz array is present. 
The \ac{HPBW} remains unchanged at 64$^\circ$ in the E-plane and 12$^\circ$ in the H-plane.
Furthermore, the results indicate that the sub-6\,GHz \ac{ULA} effectively acts as an additional ground plane for the \ac{MMW} array, causing negligible impact on the main-lobe gain while suppressing the back lobe.
Specifically, the back-lobe level decreases from $-$3.4\,dBi to $-$7.5\,dBi, thereby improving the radiation performance of the \ac{MMW} array.

\section{Conclusion} \label{sec:conclusion}
Both simulation and measurement results confirm that the presence of the \ac{MMW} array has only a minor impact on the radiation characteristics of both a single sub-6\,GHz patch element and a sub-6\,GHz \ac{ULA}.
Similarly, the sub-6\,GHz array has a negligible influence on the main-lobe performance of the \ac{MMW} array, while beneficially reducing back-lobe radiation.
These findings demonstrate that co-located sub-6\,GHz and \ac{MMW} \ac{ULA} configurations are feasible and introduce only negligible performance degradation.
Nevertheless, future work should extend this analysis to uniform planar array configurations to assess radiation effects in two-dimensional antenna structures.

\bibliography{references}
\bibliographystyle{IEEEtran}

\end{document}